\documentclass[11pt]{article}
\usepackage{amssymb}
\usepackage{amsfonts,siunitx}
\usepackage{amstext,theorem}
\usepackage{amsopn}
\usepackage{amsmath,shadow,fancybox,graphics}
\usepackage{epsf,color,url,mathtools}
\usepackage[T1]{fontenc}
\usepackage[utf8]{inputenc}
\usepackage{geometry,stmaryrd}
\usepackage[setpagesize=false]{hyperref}
\usepackage[pdflatex]{graphicx}
\usepackage{subfig}

\DeclareGraphicsExtensions{.jpg,.mps,.pdf,.png}
\DeclareMathOperator{\cotan}{cotan}

\newcommand{\qedd}{\nobreak \ifvmode \relax \else
    \ifdim\lastskip<1.5em \hskip-\lastskip
    \hskip1.5em plus0em minus0.5em \fi \nobreak
    \vrule heigh$t_0$.75em width0.5em depth0.25em\fi}

\begin{document}

\title{A 2-dimensional Geometry for Biological Time}

\author{Francis Bailly\footnote{Physics, CNRS, Meudon}, Giuseppe Longo\footnote{Informatique, CNRS – ENS and CREA, Paris, longo@di.ens.fr, \url{http://www.di.ens.fr/users/longo}} , Mael Montevil\footnote{Informatique, ENS and ED Frontières du vivant,  Paris V, Paris }}

\maketitle

\begin{abstract}
This paper\footnote{Published in   
\emph{Progress in Biophysics and Molecular Biology}, 106(3):474 – 484, 2011. doi:
\href{http://dx.doi.org/10.1016/j.pbiomolbio.2011.02.001}{10.1016/j.pbiomolbio.2011.02.001}.
} proposes an abstract mathematical frame for describing some features
of biological time. The key point is that usual physical (linear) representation
of time is insufficient, in our view, for the understanding key phenomena of
life, such as rhythms, both physical (circadian, seasonal~\dots) and properly
biological (heart beating, respiration, metabolic~\dots). In particular, the
role of biological rhythms do not seem to have any counterpart in mathematical
formalization of physical clocks, which are based on frequencies along the usual
(possibly thermodynamical, thus oriented) time. We then suggest a functional
representation of biological time by a 2-dimensional manifold as a mathematical
frame for accommodating autonomous biological rhythms. The ``visual''
representation of rhythms so obtained, in particular heart beatings, will
provide, by a few examples, hints towards possible applications of our approach
to the understanding of interspecific differences or intraspecific pathologies.
The 3-dimensional embedding space, needed for purely mathematical reasons, allows
to introduce a suitable extra-dimension for “representation time”, with a
cognitive significance.
\end{abstract}


  \tableofcontents

\section{Introduction}

{L}{iving} phenomena displays rather characteristic and specific traits; among these, manifestations of temporality and of its role are particularly remarkable: development, variegated biological rhythms, metabolic evolution, aging,~\dots. This is why we believe that any attempt at conceptualizing life phenomena --- be it only partially --- cannot avoid addressing such temporal aspects that are specific to it. In that which follows, we will examine this question from different angles in view of providing a first attempt at synthesis.

The intuitive ``geometry of time'' in physics was (and often still is) based, first, on the absolute Newtonian straight time line. This was later enriched by the order structure of Cantor type real numbers, an ordered set of points, topologically complete (dense and without gaps). Thermodynamics and the theories of irreversible dynamics (phase transitions, bifurcations, passing into chaos,~\dots) have imposed an ``arrow'' upon classical time, by adding an orientation to the topological and metric structure. But it is with relativity and quantum physics that the theorization of time has led to rather audacious reflections. In the first case, to give only one example from a very rich debate which goes so far as to introduce a circular time (proposed by Gödel as a possible solution to Einstein's equations) to Minkowski space:  by means of its famous causality cone, this space explains, within the framework of a unified geometry of space-time, the structure of any possible correlation between physical 
objects, in special relativity.

In quantum physics the situation is maybe even more complex or, in any
event, less stable. We go from essentially classical frameworks to a sometimes
two-dimensional time (in accordance with the structure of the field of complex
numbers with regard to which Hilbert spaces are defined, the theoretical loci of
quantum description), up to the audacity of Feynman’s temporal ``zigzags''
\cite{feynman1967}. This latter approach is a very interesting example of
intelligibility by means of a ``geometric'' restructuring of time:  the creation of antimatter
would cause within the $CPT$ symmetry (charge, parity, time) a symmetry breaking
in terms of charge, while leaving parity unchanged. Global symmetry is then
achieved by locally inverting the arrow of time. Another approach, with similar
motivations, is that of the fractal geometry of space-time, specific to the
``scale relativity'' proposed by \cite{nottale1993}:  in it, time is reorganized
upon a ``broken'' line (a fractal), which is continuous but non-derivable. Further interesting reflections, along similar lines, may be found in \cite{leme98}.

Physics however will remain but a methodological reference for our work,
because the analysis of the physical singularity of living phenomena 
\cite{bailly2006,bailly2011} requires a significant enrichment of the conceptual and
mathematical spaces by which we make inert matter intelligible. One of the new
features which we introduce consists in the usage that we will make of the
``compactification'' of a temporal straight line:  in short, we will try to
\emph{mathematically understand rhythms and biological cycles by means of the
addition
of ``fibers'' (a precise mathematical notion, introduced summarily below) which
are orthogonal to a physical time that remains a one-dimensional straight
line}.
From our standpoint, a living being is a true ``organizer'' of time; by its
autonomy and action, it confers it a more complex structure than the
algebraic order of real numbers, but also more than any
organization one could propose for the time of inert matter. In short, the time
of a living organism, by its specific rhythms, intimately articulates itself
with that of physics all the while preserving its autonomy. We would therefore
like to contribute to making the \emph{morphological complexity} of biological
time intelligible, by presenting a possible geometry of its structure, as a two
dimensional manifold.

The first paragraph will introduce the theme of biological rhythms. One consequence of our approach is the possibility of mathematically giving what we hope to be more precise and relevant meaning to notions that are usually treated in a rather informal fashion and unrelated between one another, such as those of representation time, physical time vs. biological rhythms,~\dots\ and this within a rigorous mathematical frame.

%This approach will enable to integrate, within the structure of the causality of living phenomena, the causal circularities which are due, in particular, to phenomena of biological integration and regulation (which we will analyze namely in terms of ``correlation lengths'' within the biological object).

\subsection{Methodological remarks}

Let’s recall that physics, in its history, was constituted according to major
dimensional constants (gravitation, the speed of light, Plank’s constant --- with
dimensions, respectively:  acceleration, speed, action). What is striking, in
biology, is the presence of a few major invariants \emph{with no dimension},
those that are specified in the rhythms of which we will speak below. The
mathematization of physics concentrated on invariants, among which the above
constants, but also those of ``objective determinations'', which we address in
length in \cite{bailly2006,bailly2011}. We suggest here to start with these rare
invariants, these constants and rhythms which are to be found in biology,
because, beyond the physico-chemical, the \emph{structural stability} of living
phenomena is not ``invariant'', physically speaking:  it is profoundly imbued with
\emph{variability}.

Observe also that in physics, time is mostly described as a parameter of the state functions describing a system. The phenomena encountered in biology, however, seem to  trigger the need of other theoretical strategies and this at many different temporal levels of organization  (physiology, ontogenesis, phylogenesis,~\dots). We will provide a  geometrical scheme of biological time that stresses the crucial role of time in life and allow to understand some of the above features mainly through the use of two theoretical concepts. 

The first one, which we will discuss in depth latter, is the ubiquity of rhythms in biological temporal organization. There is indeed  few features that are ubiquitous in biology but the iteration of similar processes seems to be one of them. We will however make a clear distinction between two type of cyclicity encountered in living systems.

The second concept is a way to understand the constitution  and
maintenance of biological organization, both in phylogenesis and embryogenesis,
that we formalized by the notion of anti-entropy in \cite{bailly2009}. That approach
allows the addition of a new theoretical aspect of time irreversibility in
biological systems, that completes and adds up to the thermodynamical
irreversibility driven by the notion of entropy. At the level both of evolution
and embryogenesis, this irreversibility manifests itself by the increase of
complexity of the organism (number of cells, number of cell types, cell networks
--- neural typically, geometrical complexity of the organs, constitution of
interacting yet differentiated levels of organization,~\dots).

Methodologically, by a duality with physics, in \cite{bailly2009} time is understood
as an operator (like energy in Quantum Physics), not as a parameter. This makes
time a fundamental observable of biology (like energy in physics) and it gives
meaning to its key role in ``biological organization'', since rhythms organize
life.

\section{An abstract schema for biological temporality.}

\subsection{Premise:  Rhythms}

We will introduce a second dimension of time,  associated to the endogenous
internal rhythms of organisms, a dimension of time which we will represent as
compacified ($\mathcal{S}_1$ topology\footnote{The circle is the
\emph{compactification} of the real number straight line, by the addition of a
point and its folding.}).

We denote this compacified time as $\theta$, which we can represent as a sort of ``circle'' with a ``radius''  $R_i$  (where  $R_i$  is the proper biological  time):  this circle expresses the temporal circularity, the iterative component, that is specific to internal rhythms.

\subsection{External and internal rhythms}

We will distinguish two types of  rhythms associated with biological organization, each referring to a distinct temporal dimension (below, we will note them as $t$ and $\theta$, respectively): 
\begin{description}
	\item[(Ext)] ``external'' rhythms, \emph{directed by} phenomena that are exterior to the organism, with a physical or physico-chemical origin and which physically impose themselves upon the organism. So these rhythms are the same for many species, independently of their size. They express themselves in terms of physical, hence dimensional, periods or frequencies (s, Hz) and the invariants are dimensional; they are described with regard to the dimension of physical time (in $\exp(\imath\omega t)$). Examples:  seasonal rhythms, the 24 hours-cycle and all their harmonics and sub-harmonics, the rhythms of chemical reactions which oscillate at a given temperature, etc.
	\item[(Int)] ``internal'' rhythms, of an endogenous origin, \emph{specific to physiological functions} of the organism, depend on purely biological functional specifications. These rhythms are characterized by periods which scale as the power $1/4$ of the organism’s mass and, when related to the life span of the organism which scales in the same way, they are expressed as pure numbers (they have no physical dimensionality). Invariants are therefore numeric. We propose to describe them with regard to a new compacified ``temporal'' dimension~$\theta$, with a  non-null radius, the numeric values then corresponding to a ``number of turns'', independently of the effective physical temporal extension (we have mentioned some examples:  heartbeats, breathings, cerebral frequencies, etc.). \label{EXT} 
\end{description}

We will now, even if we must be somewhat repetitive, describe further how these rhythms take place in biological organization, which is precisely what we would like to provide account for: 

\begin{itemize}
	\item The external rhythms (Ext) are to be identified with physical time (typically measured by a clock) determined universally --- their temporal features does not depend of the biological system we consider. Key examples include circadian, circannual or tidal cycles. The effects or the relevancy of these cycles depend of course on the organism that we consider (with possible sexual dimorphism). For example, diurnal and nocturnal animals are in phase opposition, whereas tides are mainly relevant for marine organisms, and especially in the foreshore. Whatever organism we consider, the period and the phase of these rhythms are the same as they are dependent on external physical events. In order to be a little more precise, this rhythms are generally associated with a double process:  the physical process, outside the living system (and which can be very precisely predicted) and its ``shadow'' inside the system which is kept  synchronized by so-called ``Zeitgeber'' (light for circadian cycle for example). 
This distinction leads in particular to a specific inertia, encountered for example in the  ``jet lags'' phenomenon.
	
Simple chemical   oscillations inside an organism will fall in this category too, since their period is determined by physical principles, however their phase  depends on a specific organism (a specific trajectory) since it is the organism which  constructs this   chemical system.
	
As a result, this kind of rhythms, and their subharmonics, can be considered in the usual physical way and represented by terms like $e^{\imath\omega t}$. 

	\item The second kind of rhythms, the endogenous biological cycles in
(Int), do not depend directly on external physical rhythms. They could be called
autonomous or eigen rhythms and scale with the size of the organism (frequencies
brought to a power $-1/4$ of the mass, periods brought to a power
$1/4$), which is not the case with constraining external rhythms which
impose themselves upon all (circadian rhythms, for example). Such rhythms are
encountered when we consider the heart rate, the respiratory rate, the mean life
span,~\dots, see \cite{savage2004} or \cite{lindstedt} for example. This rhythms are
naturally associated  with the number of their iterations (they can be seen as
dual variables), and these numbers provide a natural way of speaking of the age
of a biological system, yet different of the time measured by a clock. The
distinction between replicative and chronological aging for yeasts, is a clear
example of this situation, see \cite{Fabrizio04}.

It is worth noting that this kind of rhythms leads to the use and the study of pure numbers instead of quantities with a physical dimensionality (such as intervals of  physical time). Moreover these numbers seem, at least in certain cases, to have invariant\footnote{There is still some variability, but this variability appears ``naked'' when considering this numbers, whereas the mass and temperature effects come first when considering dimensional quantities.} properties, a clear and impressive example of this is the mean number of heartbeat (or respiration) during  life  which is almost invariant among mammals.
\end{itemize}	

In summary, endogenous biological rhythms: 
\begin{itemize}
	\item are  determined by pure numbers (number of breathings or heart beats over a lifetime, for example) and not, in general, by dimensional magnitudes as is the case in physics (seconds, Hertz,~\dots);
	\item  depend on the adult mass of the organism that we consider, by following the allometric law  $ \tau_i \propto W_f^{1/4}$ (for heterotherms, the temperature is involved too);
	\item in our approach, they are analyzed and put into relation to each other by adding an additional compacified ``temporal'' dimension (an angle, actually, like in a clock), in contrast to the usual physical dimension of time, a line, non-compacified and endowed with dimensionality.
\end{itemize}

 Since these endogenous rhythms co-exist with physical time, we
consider a temporality of a topological dimension equal to $2$ formed by the
\emph{direct product} of the non-compacified part, the real straight line of the
variable $t$ (the physical time parameter) and, as a fiber upon the latter, the
compacified part, a circle $\mathcal{S}_1$, of which the variable is $\theta$.
Since we consider a two-dimensional time, with a second dimension associated
with specific biological invariants, our approach is very different of the usual
approach of biological time in terms of dynamical systems, which allows to
tackle different kind of questions, like synchronization or stability (see for
example the noteworthy book of  \cite{winfree}), but do not deal with
these invariants. 

The idea of using supplementary compacified dimensions in theoretical physics has been introduced by Kaluza and Klein \cite{97}, and is still widely used in unification theories (string, superstring, M-theory,~\dots).  There are of course major differences between these uses of compacified dimensions and ours.  First they concern mostly the addition of \emph{space-}dimensions; second these dimensions are not observable in physics, whereas they are clearly observable in biology. In our approach, the projection of this second dimension on physical time leads to  quantities that have the dimension of a time; their mean follows the allometric law, as such they are parametrized by a mass (or, equivalently, by an energy; here one may see again the dual role of energy vs. time as parameter vs. operator, the duality w. r. to Quantum Physics we mentioned and extensively used in \cite{bailly2009}).

We insist that the endogenous rhythmicities and cyclicities are not physical temporal rhythms or cycles as such, as they are \emph{iterations} of which the total number is set independently from the empirical (temporally physical) life span. As we said, they are pure numbers, a few rare constants (invariants) in biology.
Our aim is that of a geometric organization of biological time which, by the generativity specific to mathematical structures, would also enable us to \emph{derive} meaning and to \emph{mathematically correlate} diverse notions.
The text itself constitutes the commentary and the specification of the following schemata, which are meant to ``visualize'' the two-dimensionality which we propose for the time specific to living phenomena. 

\section{Mathematical description}
We first consider both external and internal rhythms; later, we will mainly focus on internal rhythms of organisms (we can take as a paradigmatic example the heart rate of mammals). We begin by providing a qualitative draft of our scheme to show its geometrical structure in figure \ref{fig:schq}, then we will quantify its parameters and explain more precisely their meaning.

\subsection{Qualitative drawings of our schemata}

\begin{figure}[htbp]
\centering
\label{fig:schq-a}\includegraphics[scale=0.8]{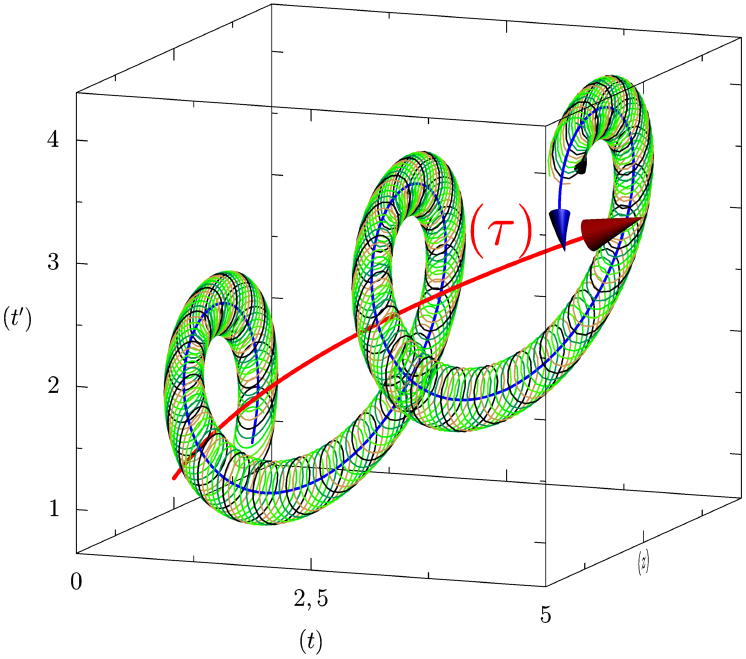}
\caption[Qualitative illustration of our geometric scheme, as a 2-dimensional manifold]{\emph{Qualitative illustration of our geometric scheme, as a 2-dimensional manifold.} In red, the global age of the organism, in blue its modulation by the physical rhythm. Here the surface is suggested by varying values of  $\theta$.}
\label{fig:schq}
\end{figure}

Following the aforementioned ideas, biological time is a (curved) surface:  thus,
it will be described in 3-dimensions (the embedding space). Note that, if we
were considering only biological rhythms, our 2-dimensional manifold would be a
cylinder:  the (oriented) line of physical time \emph{times} the extra
compacified dimension. The situation is more complicated, in view of the
further, physical rhythms we want to take into account. They do not require an extra
dimension, but they ``bend'' the cylinder, by imposing global (external)
rhythms. Thus, a proper biological rhythm is represented by what we may call a
``second order helix'', that is, a helix that is obtained (is winding) over a
cylinder, $\mathcal{C}_i$, which, in turn, is winding around a bigger cylinder, $\mathcal{C}_e$, of which the axis
is the line $(\tau)$. As basic reference, we  choose orthogonal Cartesian
coordinates. Physical time, which is oriented by thermodynamic principles of 
irreversibility  and is  measured by a clock as in classical or relativistic
physic, will be the first axis $(t)$ of our reference system and will enable the
characterization of instants and the measurement of durations. The second axis,
$(t')$, will be associated with the proper irreversibility of biological time
(for example the irreversibility of embryogenesis or, just, of ``living'', see
\ref{axis}). As such, it will represent the \emph{biological age}, or the
internal irreversible clock of the organism we consider. The $(t)$ and $(t')$
axis are oriented in the usual way ($(t)$ towards the right and $(t')$ pointing
upwards). The  third axis, $(z)$, (see \ref{fig:schq}) is generated by the
mathematical need of a 3-dimensional embedding space; yet, we claim that it has
a biological meaning that will become clear later, in section \ref{qual-draw}.

We consider the surface of the cylinder $\mathcal{C}_i $ as parametrized by $t$ (the
physical time) and $\theta \in [0, 2 \pi]$ (the compacified time).

Let's then take a further step by gradually making explicit the functional dependencies.
\begin{itemize}
\item The average progression with respect to  $(t')$ will be represented by a function $\tau\left(\frac{t-t_b}{\tau_i}\right)$. $\tau$ is a growing function due to the irreversibility of biological time, and has a decreasing derivative due to the decrease of activity during development and aging. $t_b$ is the physical time of a  biological  event of reference (time of fecundation for example). $\tau_i$ is a characteristic time of the biological activity of the adult:  for example, the mean ``beat to beat'' interval under standardized conditions (other reference systems can be chosen such as the  mean time taken to attain \SI{98}{\percent} of adult mass, life expectancy, respiratory interval,~\dots). This value represents, as a function of physical time, the  age of the system inasmuch this age is  biologically relevant (see figure \ref{fig:sch-a}:  the graph of $\tau$  lies on the $(t \times t')$ plane). Set then: 

\begin{equation}
\overrightarrow{\mathfrak{F}}_{\tau_i}(t,\theta)= \left(
\begin{array}{c}
    t \\
    \tau\left(\frac{t-t_b}{\tau_i}\right) \\
    0
\end{array}\right)
\end{equation}

\item We next consider a physical (external) rhythm of period $\tau_e$ (its pulsation is then $\omega_e=\frac{2\pi }{\tau_e}$) that affects the activity rate of the organism 
--- the circadian rhythm,  for example, leads to $\tau_e=24$ hours. This produces a winding spiral or helix, $ \mathcal{C}_e$ (see figure  \ref{fig:sch-b}:  here we need the third dimension $(z)$ for the embedding space of our manifold). In the definition  of $\overrightarrow{\mathfrak{G}}_{\tau_i}(t,\theta)$, $R_e$ represents the impact of this physical rhythm on biological activity: 

\begin{equation}\label{core}
\overrightarrow{\mathfrak{G}}_{\tau_i}(t,\theta)=\overrightarrow{\mathfrak{F}}_{\tau_i}(t,\theta) + \left(
\begin{array}{c}
    0 \\
     \frac{R_e }{\omega_e \tau_i} \cos\left( \omega_e t\right)  \\
    \frac{R_e }{\omega_e \tau_i} \sin\left( \omega_e t\right)
\end{array}\right)
\end{equation}
The term $\frac{1}{\omega_e \tau_i}$ is proportional to the number of iterations of the compacified time during one period of the physical rhythm, as such it can be considered as the temporal weight of this rhythm for an  organism (mean number of heartbeat during a day, for example), it allows to understand that a year is more important for a mouse and for an elephant. As a consequence the radius of $ \mathcal{C}_e$ is proportional to both the impact $R_e$ of this rhythm on biological activity, and on the weight of this rhythm in terms of number of iteration of the endogenous rhythm considered during one period of the physical rhythm. 

\begin{figure}[htbp]
\centering
  \subfloat[]  { \label{fig:sch-a}\includegraphics[scale=0.85]{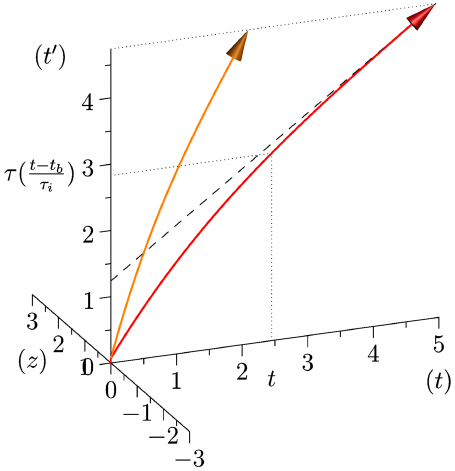}}\hfill
  \subfloat[]{ \label{fig:sch-b}\includegraphics[scale=0.85]{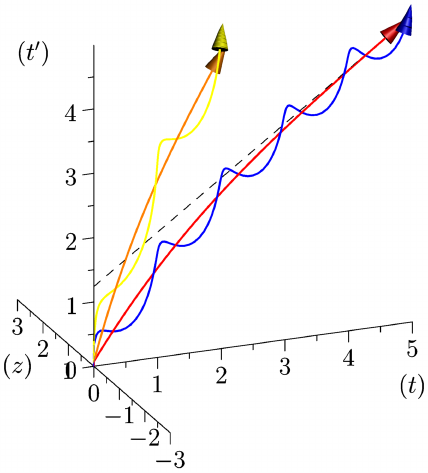}}
 \caption[Qualitative Illustration of the first components of our model]{\emph{Qualitative Illustration of the first components of our model.}  \textsc{Left}, the function $\tau\left(\frac{t-t_b}{\tau_i}\right)$, which represents the global age of an organism:   this age increases at a greater pace during development and slows down progressively, see section \ref{axis}. In orange a small mammal (a mouse for example) and in red a bigger one (an elephant). The life span of the first is shorter than one of the second. \textsc{Right}, in blue (and yellow), a physical rhythm has been added (this rhythm is very slow for illustrative purposes). Notice that this physical rhythm has the same period for both animals, but one of its iteration has a greater weight for the smaller animal.}
 \label{fig:sch}
\end{figure}

\item We can finally add a biological (internal) rhythm, which depends on an increasing function $s_{\tau_i}(t)$ (see figure \ref{fig:scen}). $s_{\tau_i}(t)$ has a proper biological meaning:  for example, if we impose $s_{\tau_i}(t_b)=0$, with $t=t_b$ when the heart begins to beat\footnote{Let's remark that, unlike in physics --- classical, relativistic or quantum--- biological time has an origin, whatever level of organization we consider. As a result there is no time-symmetry for translations, a fundamental property, in (relativistic) physics for the constitution of invariants, e.g. energy conservation.}, $s_{\tau_i}(t)$   is the number of heartbeats of the organism at time $t$, and thus the mean maximum of $s$, obtained when death occurs, does not depend on the  organism we consider (among mammals, typically). Set then, for $\overrightarrow{\mathfrak{G}}_{\tau_i}(t,\theta)$ as in equation \ref{core}: 

\begin{equation}\label{phyry}
\overrightarrow{\mathfrak{T}}_{\tau_i}(t,\theta)=\overrightarrow{\mathfrak{G}}_{\tau_i}(t,\theta)+ \left(
\begin{array}{c}
    0\\
    R_i\cos\left(2 \pi s_{\tau_i}(t)+\theta\right) \\
     R_i \sin\left(2 \pi s_{\tau_i}(t)+\theta\right)
\end{array}\right)
\end{equation}

\end{itemize}

\subsection{Quantitative scheme of biological time}
\label{qual-draw}

  Now, one of  the simplest way to define more precisely $s$ is to use $\overrightarrow{\mathfrak{G}}_{\tau_i}$ and more precisely the length of the curve defined by $\overrightarrow{\mathfrak{G}}_{\tau_i}$. We obtain then for the instantaneous pulsation, where $\tau'$ is the derivative of $\tau$ (thus $\frac{d}{dt}\tau\left(\frac{t-t_b}{\tau_i}\right)= \frac{1}{\tau_i}\tau'\left(\frac{t-t_b}{\tau_i}\right)$) and the other components are the derivative of the remaining coordinates in equation \ref{phyry}: 

\begin{equation}\label{metric0}
\frac{d s_{\tau_i}(t)} {dt}=
\sqrt{ \alpha^2 \times 1^2  +\left(\frac{\tau'\left(\frac{t-t_b}{\tau_i}\right)}{\tau_i}
-\frac{R_e }{ \tau_i} \sin\left( \omega_e t\right) \right)^2 +\left(\frac{R_e }{ \tau_i}\right)^2
\cos\left( \omega_e t\right)^2}
\end{equation}

The term $\alpha$ is here for (physical) dimensionality reasons:  since  our metric has the
dimension of a frequency, and $\frac{dt}{dt} = 1$, the derivative of the first
component of the vector in equation \ref{core}, has no dimension, then we need
to introduce this coefficient whose dimension is a frequency\footnote{This kind
of reasoning is commonplace in physics.}. 

When $\alpha = 0$ we can simplify \ref{metric0} to: 

\begin{equation}\label{metric}
\frac{d s_{\tau_i}(t)} {dt}=
\sqrt{ \left(\frac{\tau'\left(\frac{t-t_b}{\tau_i}\right)}{\tau_i}\right)^2
+\left(\frac{R_e }{ \tau_i}\right)^2  - 2\frac{R_e
\tau'\left(\frac{t-t_b}{\tau_i}\right)}{\tau_i^2} \sin\left( \omega_e t\right)}
\end{equation}

Now, if we consider hibernating
animals, or frozen organisms, we have situations where the physical time flows
normally but where the biological time  almost stops or even totally stops. For
$\alpha\neq 0$, even in the frozen case, biological time would flow with
$\frac{d} {dt} s_{\tau_i}(t)\geq \alpha$.  It
seems then natural to suggest that $\alpha=0$. Moreover, for $\alpha=0$, we go
back to allometric relations, since, in this case, $\frac{d}
{dt} s_{\tau_i}(t)$ is
proportional to $\frac{1}{\tau_i}$. Now, $\tau_i$ is proportional to
$W_f^{1/4}$, by allometry, and, thus, $\frac{d} {dt} s_{\tau_i}(t)$,
which is a frequency, to $W_f^{-1/4}$, as it should be.

Another way to express this is to say that physical time \emph{per se}
does not make biological organization get older:  it is only when there is a
biological activity (which in return is of course always associated with
physical time) that aging appears.

We can now even give a meaning to the third, $(z)$, axis:  since
$\tau\left(\frac{t-t_b}{\tau_i}\right)$ is on the $(t\times t')$ plane, a positive $(z)$
corresponds to a positive $ \sin\left( \omega_e t\right)$, by equation \ref{phyry}, and it
is  associated with a slowdown of biological activity (sleep, for example),
whereas the negative values are associated to a faster pace (wake for example).

As a fundamental feature of the model that we will analyze next, we assume that the speed of rotation with respect to the compacified time is constant, which leads to a  radius $R_i=\text{Cst}$.
\begin{equation}
\left\|\frac{\partial \overrightarrow{\mathfrak{T}}_{\tau_i}(t,\theta)}{ \partial \theta}\right\|=R_i(t,\theta)=\text{Cst}
\end{equation}

This assumption  ``geometrizes'' time even further:  acceleration and slow-down will be \emph{seen} as contraction and enlargement of a cylinder in \S \ref{varia}. In that section, as an application, we will develop a geometrical analysis of biological rate variability, and, as an example, we will consider heart rate. Note that this radius $R_i$ is the dimension accommodating the biological rhythms, thus it is not a physical dimension (it is a pure number). Our assumption is consistent with the idea that each iteration along the compacified time contributes equally to aging.

\section{Analysis of the model}

In this section we will explore the various biological aspects our approach allows to put together, mainly on the questions of interspecific and intraspecific  allometry and on (heart) rate variability.

\subsection{ Physical periodicity of compacified time}

Since $\frac{d} {dt} s_{\tau_i}(t)$ provides the frequency of the biological rhythm, it is interesting to look for a simple analytical expression of the period associated.   To do so,  we perform a Taylor development (under the hypothesis $\tau'\left(\frac{t-t_b}{\tau_i}\right)\gg R_e$) of the inverse of equation \ref{metric}, and as a result we obtain an approximation of the physical time associated with an iteration of the compacified time (the time between two heartbeats for example): 
\begin{equation}
\frac{1}{\frac{d s_{\tau_i}(t)} {dt}} \simeq \tau_i \left( \frac{1}{\tau'\left(\frac{t-t_b}{\tau_i}\right)}+  \frac{R_e } {\tau'\left(\frac{t-t_b}{\tau_i}\right)^2} \sin\left( \omega_e t\right) \right)
\end{equation}

We can observe several things here.
First, for adults (i.e.:  $\tau'\left(\frac{t-t_b}{\tau_i}\right)\simeq Cst$ and it does not depend on the size of the organism we consider) the result has the form  $\tau_i \left( a+  b \sin\left( \omega_e t\right) \right)$. As a consequence, when we take different species,  there is no variation of the ratio ($\frac{\tau_i b}{\tau_i  a}$) between the continuous and the period components of the biological rates. Alternatively the ratio between the rates of the biological rhythm during the slow period of the physical rhythm (sleep for example) and during  the fast period  (wake) does not depend of the species either. This result holds experimentally (see  for example \cite{savage2004} and \cite{mortola2004}). 

On the other side,  the relationship between this two rates is not linear in intraspecific variations (i.e.:  when $\tau'$ is not constant, mainly during development), and the variation of the coefficient of the rhythmic  component ${R_e } {\tau'\left(\frac{t-t_b}{\tau_i}\right)^{-2}}  $  is far greater than that of the steady (continuous) component $ {\tau'\left(\frac{t-t_b}{\tau_i}\right)^{-1}}$. This mathematical deduction agrees with experimental results, since  \cite{massin2000}, for example, find that the continuous component varies like $t^{0.16}$ while the sinusoidal part (associated with the circadian rhythm ) varies like $t^{0.75}$ for humans (between 2 months and 15 years).

\subsection{Biological irreversibility}
\begin{figure}[htbp]
\begin{minipage}{\columnwidth}\hfill
  \subfloat[Cofluent case]{\label{fig:scen-a}\includegraphics[scale=0.27]{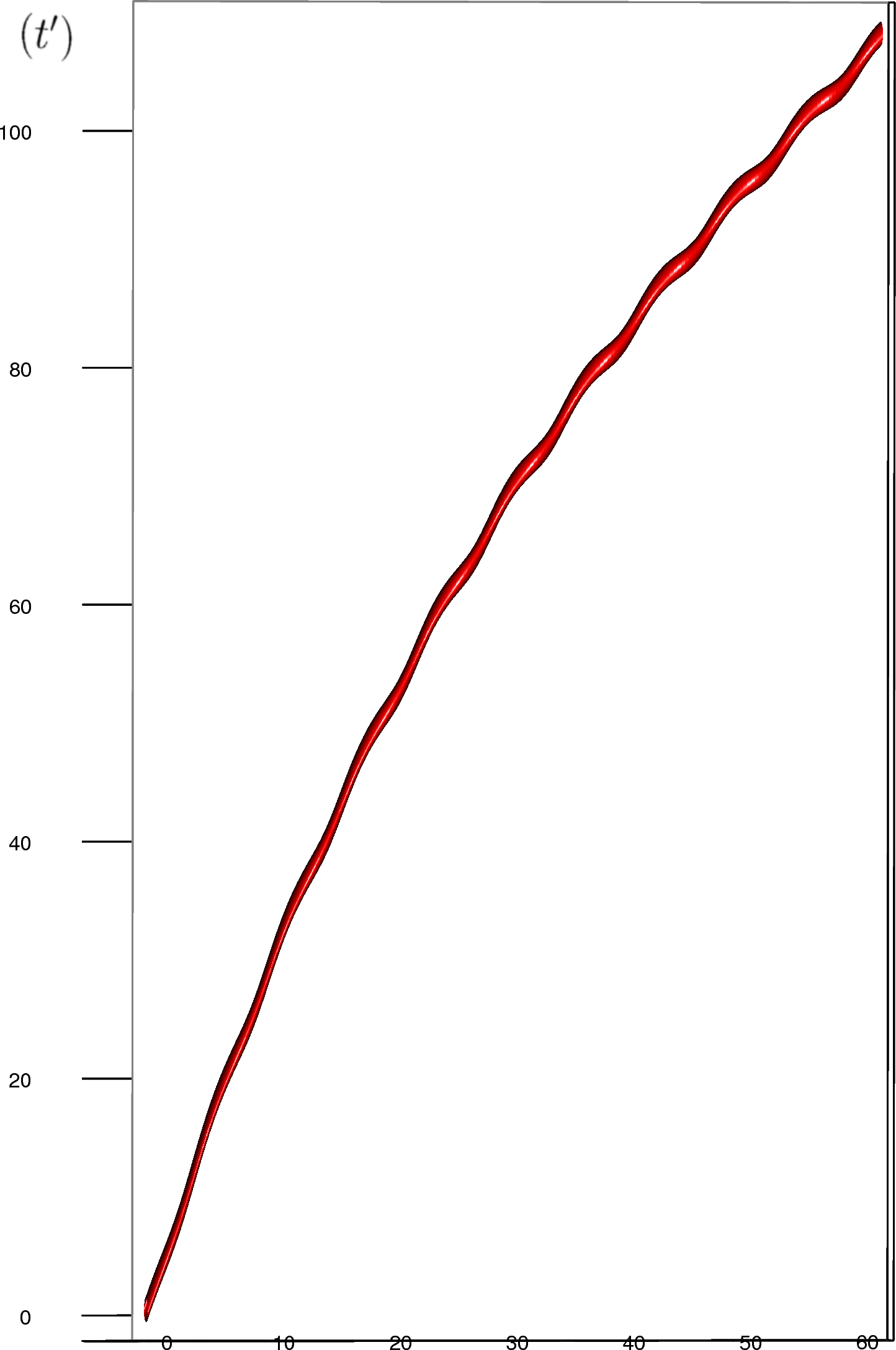}}\hfill
  \subfloat[Minimally cofluent case]{\label{fig:scen-b}\includegraphics[scale=0.27]{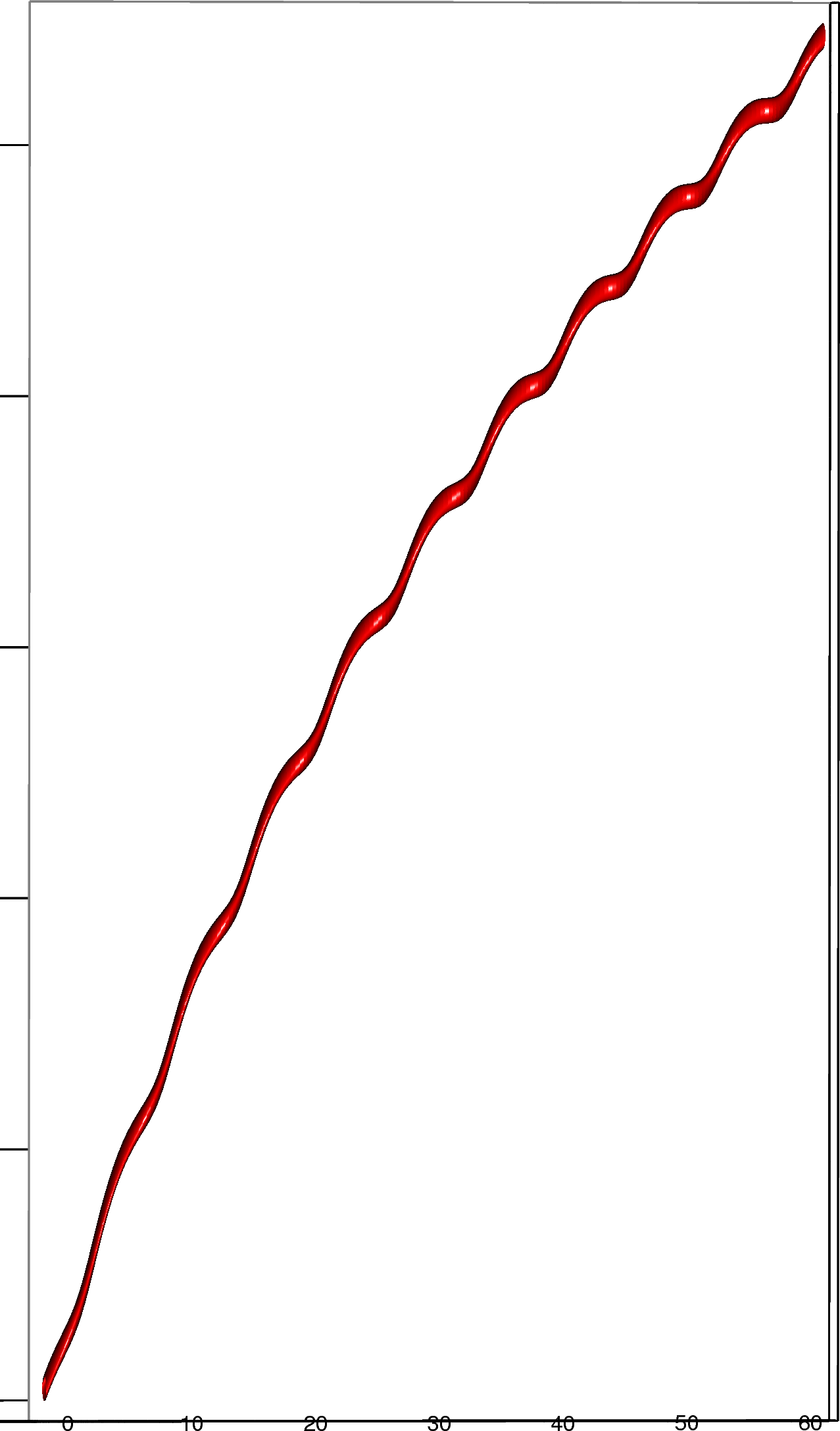}}\hfill
 \subfloat[Non-cofluent case]{\label{fig:scen-c}\includegraphics[scale=0.27]{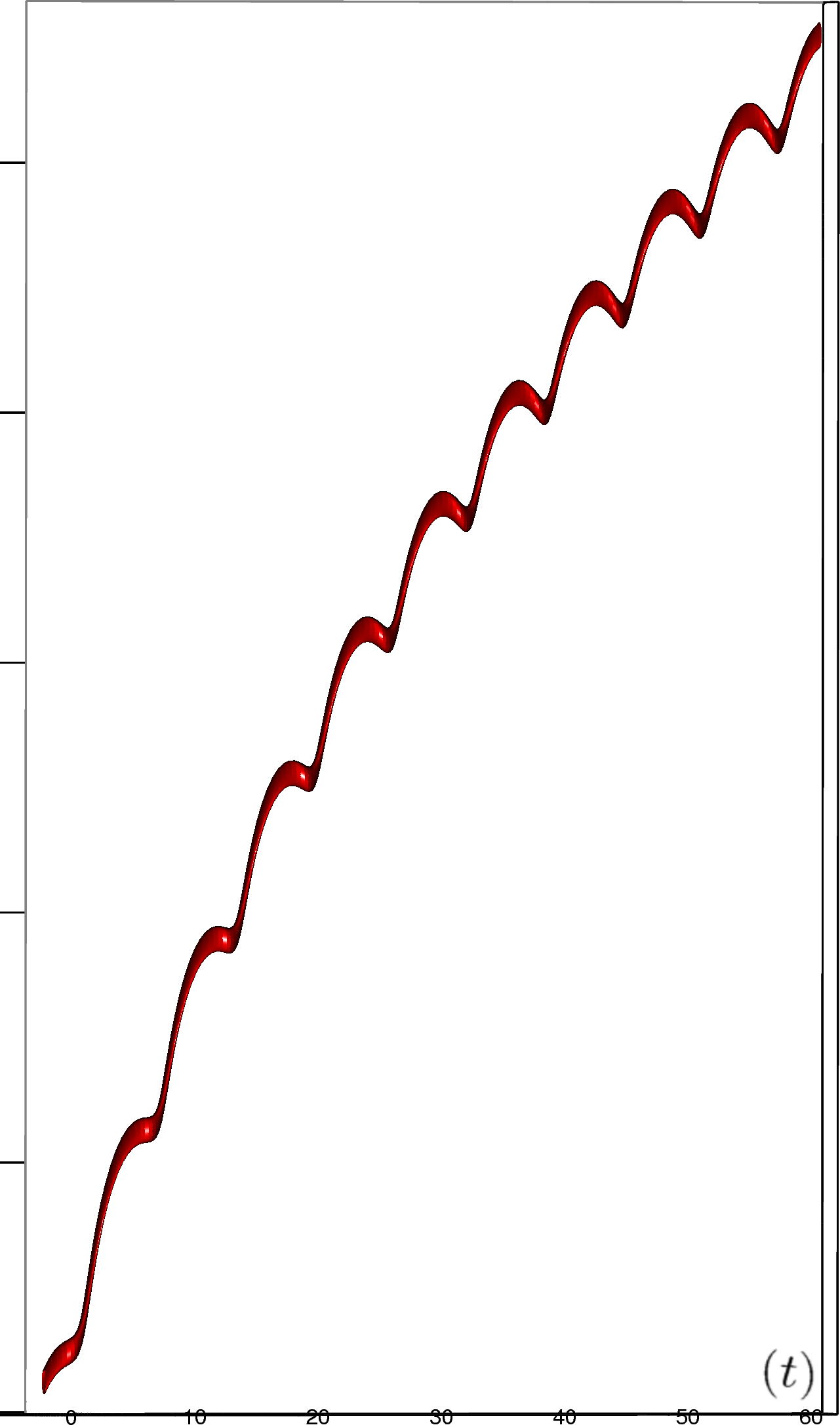}}
\end{minipage}
\begin{minipage}{\columnwidth}
   \subfloat[Cofluent case]{\label{fig:scen-d}\includegraphics[scale=0.14]{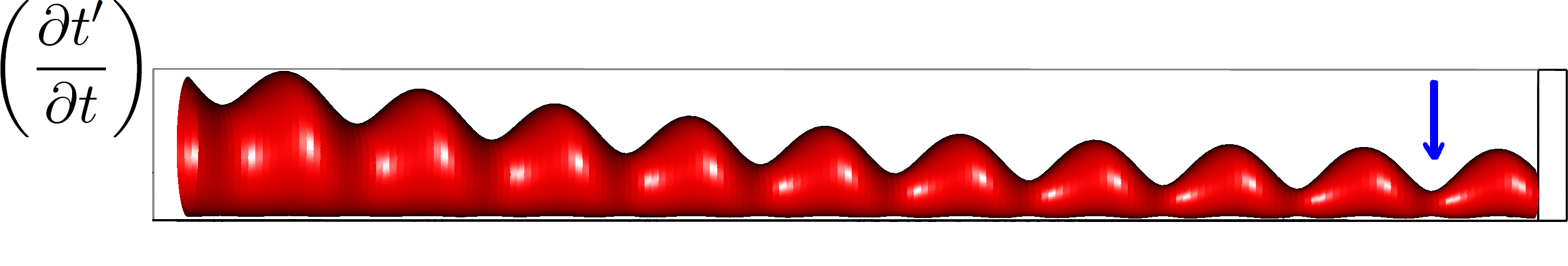}}\hfill  \subfloat[Minimally cofluent case]{\label{fig:scen-e}\includegraphics[scale=0.14]{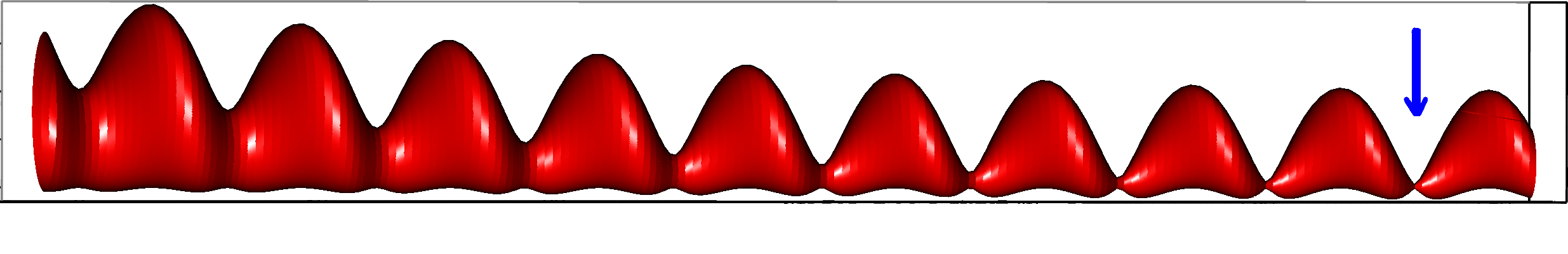}} \hfill \subfloat[Non-cofluent case]{\label{fig:scen-f}\includegraphics[scale=0.14]{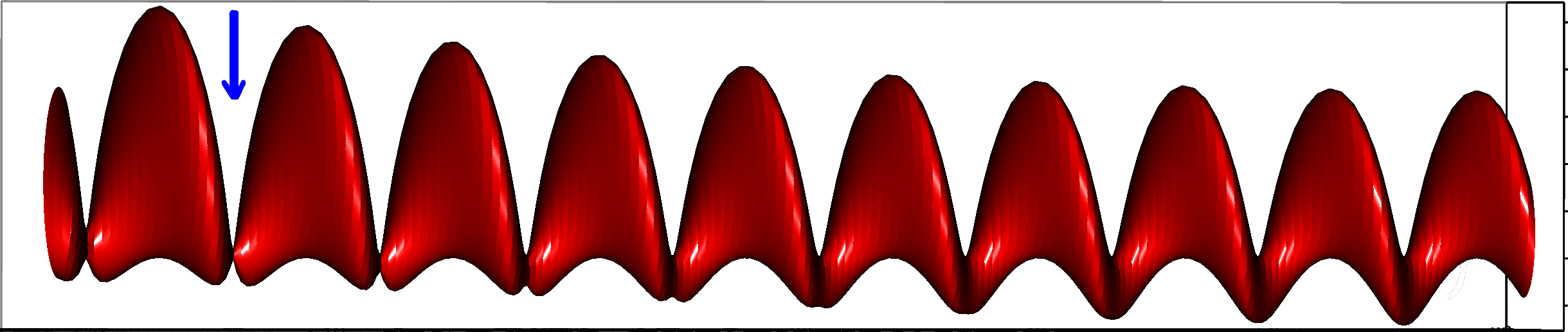}}
\end{minipage}
 \caption[Illustration of the three scenari]{\emph{Illustration of the three scenari.} \textsc{Top}:  the scheme $\overrightarrow{\mathfrak{T}}_{\tau_i}(t,\theta)$ and \textsc{bottom} its time derivative $\frac{\partial \overrightarrow{\mathfrak{T}}_{\tau_i}(t,\theta)}{ \partial t}$.  \textsc{From left to right}:  Cofluent case, minimally cofluent case and non-cofluent case. Since the radius of the compacified time is proportional to its physical rate when looking at $\frac{\partial \overrightarrow{\mathfrak{T}}_{\tau_i}(t,\theta)}{ \partial t}$ (see \S \ref{radiusfreq}), the bottom pictures allows to see when the slowest rate occurs (i.e.:  when the radius is the smallest, blue arrow. Here respectively:  for adults in figure  \ref{fig:scen-d} and \ref{fig:scen-e} and for infants in \ref{fig:scen-f}).}\label{fig:scen}
 \end{figure}
We can now look more precisely at the second axis, $(t')$, of our reference system. Since this aspect of biological time is irreversible and flows in the same direction than physical time ($\tau(t)$ is an increasing function of $t$),  $\overrightarrow{\mathfrak{G}}_{\tau_i}$ in equation \ref{core} should increase with respect to this direction. When this condition is met, we will  say that these times are ``\emph{cofluent}''. This can be easily mathematized by looking at the partial derivative of the $(t')$ component of $\overrightarrow{\mathfrak{G}}_{\tau_i}$ (obtained with the dot product by the unitary vector $\overrightarrow{e}_{t'}$) with respect to $t$: 

\begin{equation}
\frac{\partial \overrightarrow{\mathfrak{G}}_{\tau_i}(t,\theta)}{ \partial t}.\overrightarrow{e}_{t'}=\frac{1}{\tau_i} f'\left(\frac{t-t_b}{\tau_i}\right) - \frac{R_e }{ \tau_i} \sin\left( \omega_e t\right)
\end{equation}

We obtain then three different scenari, assuming that  $\tau'\left(\frac{t-t_b}{\tau_i}\right)$ tends to be a constant for adults (and seniors), written $\tau'\left(\frac{t_\infty}{\tau_i}\right)$. We then use equation \ref{metric} to derive their \emph{observable} consequences: 

\begin{description}
\item [$ \tau'\left(\frac{t_\infty}{\tau_i}\right) >  R_e $.] In this case, biological age and the physical clock are cofluent, and the minimum rate is achieved during adult  sleep (figure \ref{fig:scen-a} and  \ref{fig:scen-d}).

\item  [$\tau'\left(\frac{t_\infty}{\tau_i}\right) \simeq  R_e $.] In this case, they are minimally cofluent, the derivative tends to zero (during night or winter) when the organism grows older, that is the rate of the biological rhythm tends to $0$ during  the   (physical) time of little biological activity.  It seems to be particularly relevant for hibernation (figure \ref{fig:scen-b} and  \ref{fig:scen-e})\dots.

\item  [$  \tau'\left(\frac{t_\infty}{\tau_i}\right) <  R_e  $.] in this case they are no longer cofluent, the nullification of the biological rate would appear during development, and, as a result, the slowest biological rhythm would appear during sleep of young individuals (figure \ref{fig:scen-c} and  \ref{fig:scen-f}).
\end{description}

This  case analysis has an actual correspondence with empirical data  for the first two cases (see for example \cite{hellbrugge1964circadian,cranford1983body}). We believe that theoretically biological time should be always cofluent so that the third case should never be realized. Indeed, the existing data,  which are mostly given for humans, confirm that case $3$ does not hold (young individuals have slow rhythms, during sleep typically, which are faster than adults slow rhythms).

It would be nice that our theoretical deduction, excluding, like in physical reasoning, the third mathematical possibility as meaningless, were empirically confirmed in large phyla. Conversely it would be also interesting if  this theoretical derivation leads to the discovery of species where also the third case is realized.   

\subsection{Allometry and physical rhythms}

When we consider organisms with different adult masses ($W_f$), we obtain a variation of $\tau_i$ according to the scaling relationships ($\tau_i \propto W_f^{1/4 }$), whereas $\omega_e$ does not change. As a result, this change corresponds to a dilatation of the $(t)$ axis (as far as $f$ is concerned) whereas the physical rhythm  modifies the geometry of biological time because the variations it triggers are anchored to the physical value $\omega_e$ (see figures  \ref{fig:sch-a}, without physical rhythm, and  \ref{fig:sch-b}, with physical rhythms.). 

Then, it is the interplay between physical rhythms and biological ones that breaks the symmetry (by dilatation) between organisms of different (adult) masses that have the same temporal invariants (most mammals for example). As a result, in this situation, the  physical conditions can be seen as constraints or frictions on biological temporal organization. Our point of view can be compared to the dimensionless time in \cite{West2005}, but they only consider the autonomous aspect of biological time, thus not considering this important interplay.   

An other way to illustrate these aspects is to count the lifelong number of iterations of cycles:  for  biological cycles, this number does not vary much when considering different species, whereas  it is strictly proportional to life span for physical ones.

\subsection{Rate Variability}
\label{sec:hrv}
Let us first introduce informally the applications we will hint to in this section, where the data are obtained from the medical references in place. Our approach to biological time leads naturally, as we will further specify, to a representation by a cylinder whose radius is proportional to the cardiac \emph{rate}. If we assume that n heartbeats yield a complete rotation around the cylinder, then a faster heart rate would appear as a circular outgrowth (a sudden increase in the radius). In this representation, a healthy individual has a complex cardiac dynamics during the day, with frequent rhythms' accelerations of varying length (from seconds to many hours). This shows up in the figures by the many circular outgrowths of different radii. On the contrary, an individual with an artificially regulated pace (with a pacemaker, say) gives a relatively smooth cylinder. The last figure below corresponds to a sudden cardiac death, without particular symptoms.

Of course we do not provide a \emph{theoretical determination} of spontaneous biological rates variability, but just a \emph{geometrical representation}. As a matter of fact in our framework, it is  quite straightforward to explore the    \emph{structure} of   biological rhythms and of their variations. More precisely, we can easily and effectively represent raw datas (for example the series of ``beat to beat'' interval over time). As a result,   we obtain   more than a qualitative schema:  it is a theoretical grounded representation of the ``anatomy'' (and pathological anatomy) of biological time.
First we need to see how we can use scales in our framework.

\subsubsection{Renormalization}

If we want to consider $n$ iterations of the compacified time $\theta$ as an iteration of an other compacified time $\tilde{\theta}$ we obtain $ \tilde{\theta}=\frac{\theta}{n}$ and $\tilde{s}_{\tau_i}=\frac{s_{\tau_i}}{n}$,  then by some sort of renormalization using the principle of constant speed for the compacified time, one has: 
\begin{equation}
\left\|\frac{\partial \overrightarrow{\mathfrak{T}}_{\tau_i}(t,\theta)}{ \partial \theta}\right\|=\frac{\tilde{ R}_i}{n}=\text{Cst}
\end{equation}
So $\tilde{R}_i= R_i n$. This result is exactly (modulo a global dilatation of the  $(t')$ and $(z)$ axis by a factor $\frac{1}{n}$) what we obtain if we construct directly our system at the level of $n$ iterations.

\label{radiusfreq}
\begin{figure}[htbp]
\centering 
   \includegraphics[scale=0.8]{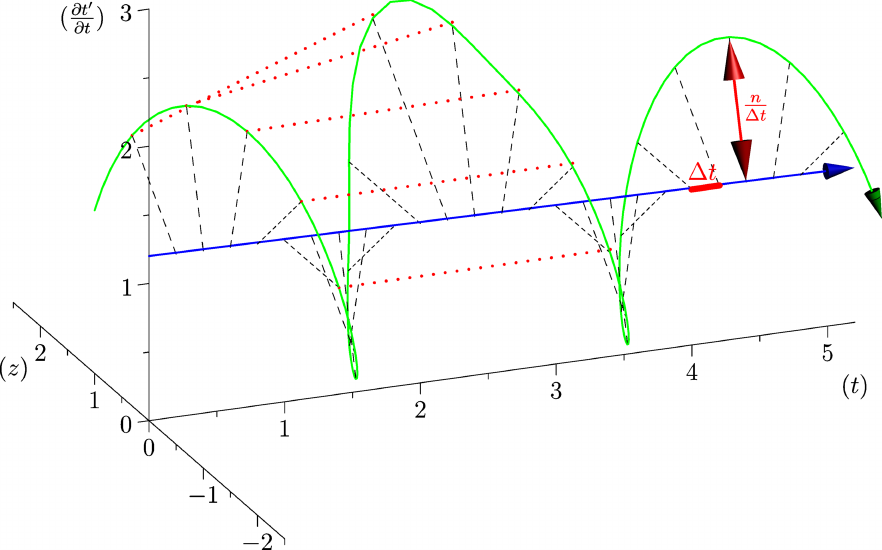}
%{\includemovie[3Dlights=Headlamp,text={\includegraphics[scale=0.7]{exp-4+0_0.pdf}},label=,toolbar=true,3Daac=0,3Dc2c=0 0 1,3Dcoo=0 0 -9240,3Droll=0,3Droo=9240,3Dbg=1 1 1,3Djscript=exp-4+0.js]{426pt}{266pt}{exp-4+0.prc}}
 \caption[Renormalization and principles of variability representation]{\emph{Renormalization and principles of variability representation.} Here, we consider $\frac{\partial \overrightarrow{\mathfrak{T}}_{\tau_i}(t,\theta)}{ \partial t}$ and we renormalize the compacified time by $n=10$. A change of speed for the  iteration $m$ of the original  compacified time appears as a sharp contrast between this iteration and its neighbors:  iteration $m-1$, $m+1$, $m-10$, $m+10$.  As a result, if there is a coherence for 10 successive iterations, we obtain a fully circular outgrowth or contraction (for an acceleration or a slowdown respectively).}
\label{fig:ren}\end{figure}

\subsubsection{Rate Variability}
\label{varia}
\begin{figure}[tbp]
%\newgeometry{left=0.9in,right=0.8in}
  \begin{center}
  \subfloat[A global view (2 days).]  {\includegraphics[scale=0.373]{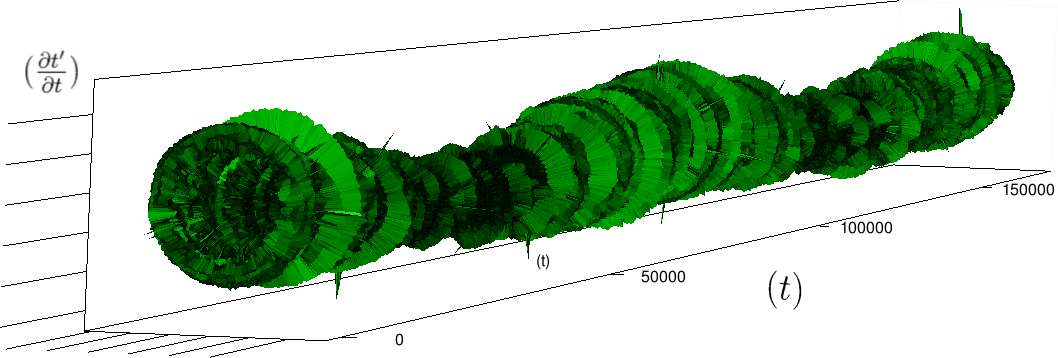}}\\
    \subfloat[Night, groups of 200 beats]    {\label{fig:scales-a}\includegraphics[scale=0.16]{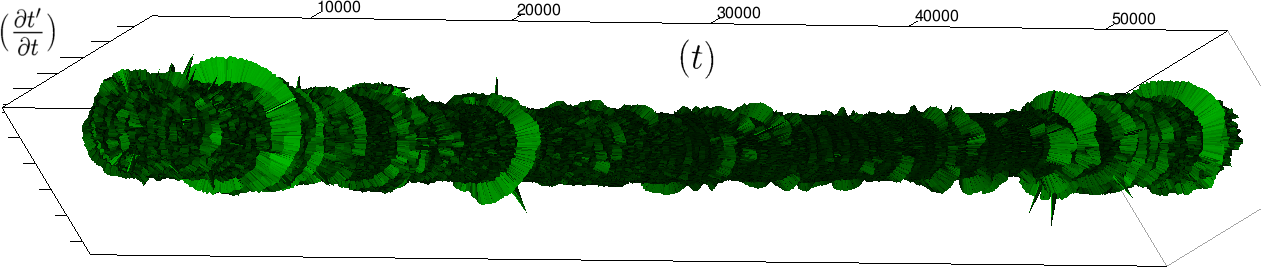}}\hfill
    \subfloat[Day, groups of 200 beats]    {\label{fig:scales-b}\includegraphics[scale=0.16]{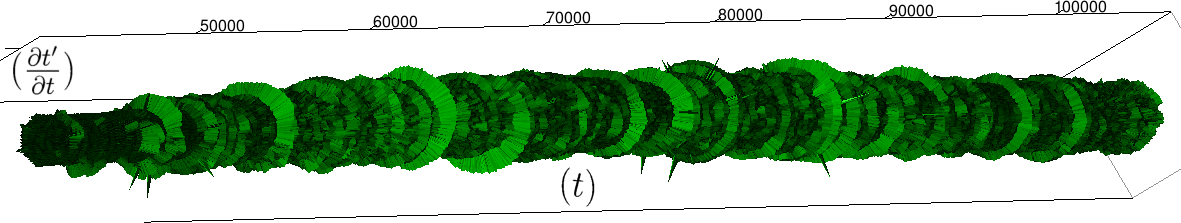}} \\
    \subfloat[Night, groups of 600 beats] {\label{fig:scales-c}\includegraphics[scale=0.155]{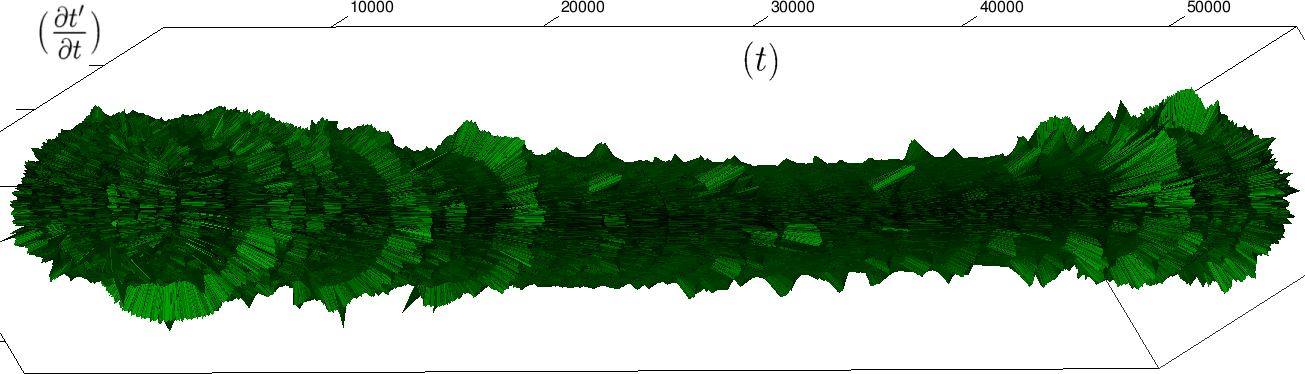}}\hfill
        \subfloat[Day, groups of 600 beats] {\label{fig:scales-d}\includegraphics[scale=0.155]{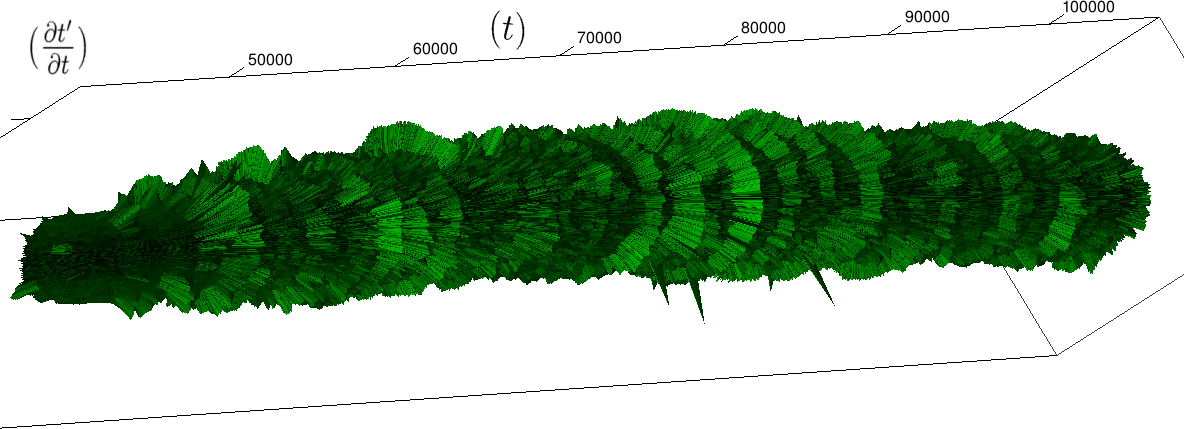}}
  \end{center}
  \caption[Comparison of the situations during sleep and wake]{\emph{Comparison of the situations during sleep and wake.} The point to notice here, is that the structure tends to become a regular cylinder during night at high scales, whereas the  wake is always complex. (Sample s20011 from The Long-Term ST Database, \cite{PhysioNet}). The series of beat to beat intervals provided by this database is used  directly, in our framework, to estimate the few parameters we need and more importantly to provide the radii involved (each heartbeat is represented).}
  \label{fig:scales}
 % \restoregeometry
\end{figure}

 If we look at the function obtained by taking the derivative of $\overrightarrow{\mathfrak{T}}_{\tau_i}(t,\theta)$ with respect to $t$, we obtain: 
\begin{equation}
 \frac{\partial \overrightarrow{\mathfrak{T}}_{\tau_i}(t,\theta)}{ \partial t}=\left(
\begin{array}{c}
    t \\
    \frac{1}{\tau_i}\tau'\left(\frac{t-t_b}{\tau_i}\right) -\frac{R_e }{ \tau_i} \sin\left( \omega_e t\right) - 2\pi R_i s'_{\tau_i}(t) \sin\left(2 \pi s_{\tau_i}(t)+\theta\right) \\
  \frac{R_e }{ \tau_i} \cos\left( \omega_e t\right)+2\pi  R_i s'_{\tau_i}(t) \cos\left(2\pi s_{\tau_i}(t)+\theta\right)
\end{array}\right)
\end{equation}
Here, instantaneous heart rate, $2\pi s'_{\tau_i}(t)$, appears directly as the radius of compacified time, (which has the physical dimension of a frequency now).

If the experimental time of each heartbeat is given in a list $(t(m))_{1\leq m \leq M}$,  we obtain a discrete empirical version of $\frac{\partial \overrightarrow{\mathfrak{T}}_{\tau_i}(t,\theta)}{ \partial t}$, renormalized by $n$: 
\begin{equation}
 \frac{\partial \overrightarrow{\hat{\mathfrak{T}}}_{\tau_i}(m)}{ \partial t}=\left(
\begin{array}{c}
    t(m) \\
    \hat{A} -\hat{R} \sin\left( \omega_e t(m)\right) - 2\pi \frac{n}{t(m+1)-t(m)} \sin\left(\frac{2 \pi m}{n} \right) \\
  \hat{R} \cos\left( \omega_e t(m)\right)+2\pi \frac{n}{t(m+1)-t(m)} \cos\left(\frac{2 \pi m}{n} \right)
\end{array}\right)
\end{equation}
where $\hat{A}$ is an estimation of $\frac{n}{\tau_i}\tau'\left(\frac{t-t_b}{\tau_i}\right)$ which may be soundly considered constant during the few days of the measure.  $ \hat{R}$ is an estimation of $n\frac{R_e }{ \tau_i}$. Both of these values are estimated by using equation \ref{metric} and $(t(m))_{1\leq m \leq M}$. We obtain a 2-dimensional structure by using triangles between adjacent points, that is to say for  $m\leq M-n-1$, the triangles  $( m, m+1, m+n )$ and $(m,m+n,m+n+1)$.
It is worth mentioning that this approach allows to obtain an empirical version of $ \overrightarrow{\mathfrak{T}}_{\tau_i}(t,\theta)$ too.

 The renormalization by $n$ allows to observe directly the correlations between $n$ consecutive heartbeats (a full circle) and the contrasts between a group and its neighbors (see figure \ref{fig:ren}), thus discriminating easily between the sleep situation (no correlations wider than $\simeq 100$ heart beat) and the healthy  wake state (correlations at each scale). The latter is indeed characterized by a succession of randomly spaced outer circle (see figure \ref{fig:scales}).

\begin{figure}[htbp]
\centering
\includegraphics[scale=0.42]{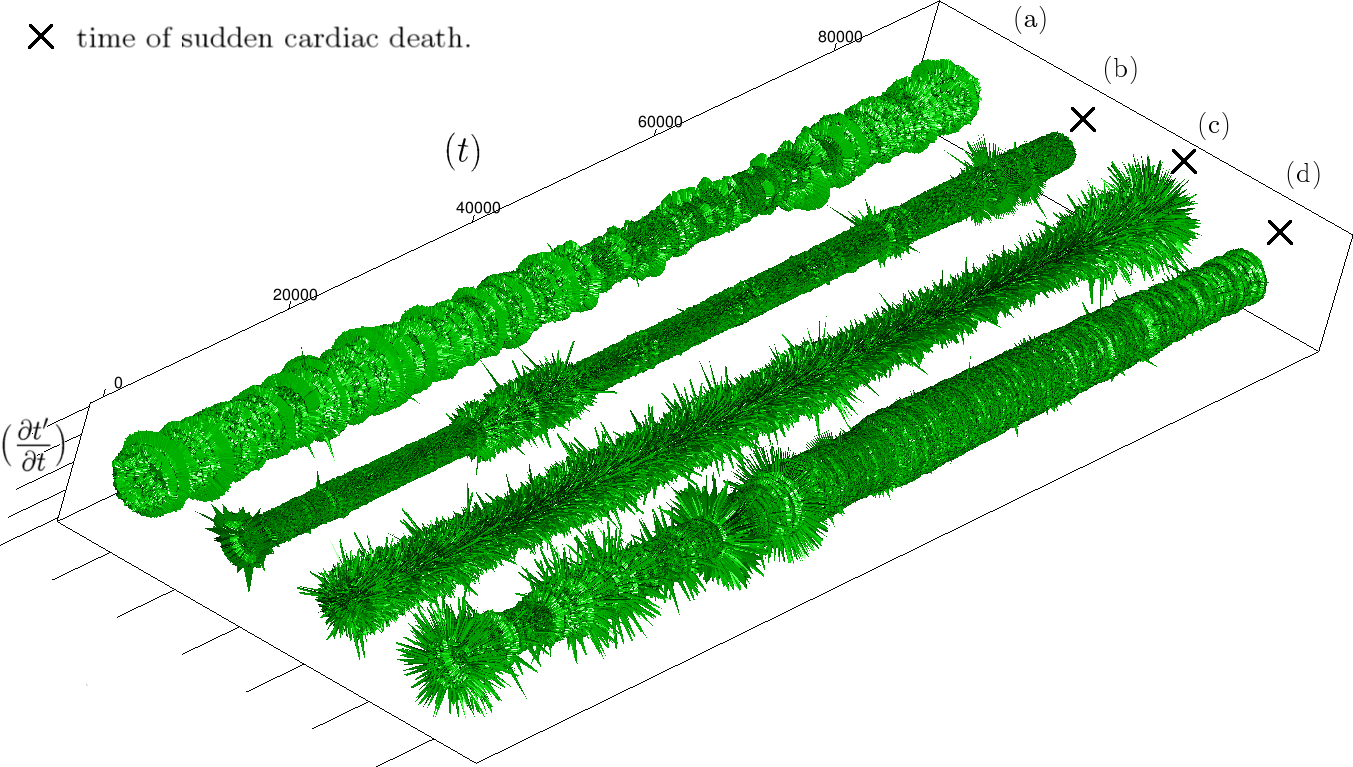}
\caption[Comparison between a healthy situation and cases of sudden cardiac arrest]{\emph{Comparison between a healthy situation and cases of sudden cardiac arrest.} (a) Healthy case, cf figure \ref{fig:scales}. (b)  Female  aged 67 with     sinus rhythm and intermittent pacing. (c) Female, 72,     with   atrial fibrillation. (d)  Male, 43,  with sinus rhythm. (The data are from samples 51, 35 and 30 from The Sudden Cardiac Death Holter Database, see \cite{PhysioNet}). }
 \label{fig:path}
 \end{figure}

Moreover this representation may be useful to study cases of heart diseases and even aging, since  this situations are characterized by an alteration of heart rate variability. We illustrate this alteration in cases of sudden cardiac death in figure \ref{fig:path} computed with datas from the The Sudden Cardiac Death Holter Database, see \cite{PhysioNet}. This figure evidentiate  the anatomy and the pathological anatomy of heart rhythms and suggests the extension of this approach to other biological rhythms which are less explored. 
\begin{itemize}
  \item  Figure \ref{fig:path}a is an example of a healthy case, which is characterized by a complex temporality during wake.
 \item In  figure \ref{fig:path}b, (intermittent) pacing leads to an excessively regular cylinder, with few heart rate variability.
\item Atrial fibrillation  in the figure \ref{fig:path}c (a kind of arrhythmia, see comments in figure \ref{fig:path}) leads to an ``hairy'' structure, which represents  a strong short term randomness (few correlations between successive heartbeats).
\item Last but not least the figure \ref{fig:path}d is not associated with a specific diagnosis (put aside sudden cardiac death at time $9000$) but it clearly shows  a very simpler  structure than the  healthy case. 
\end{itemize}

  Our approach allows to discriminate all these various cases by rather striking geometrical differences. Wavelet analysis is often used for the same purpose, but this approach is based on a massive reorganization of datas,  through a decomposition in various components, whereas we only perform a geometrical and synthetic composition of them.

\section{More discussion on the general schema \ref{fig:schq}.}
\label{axis}

\subsection{The evolutionary axis \texorpdfstring{$(\tau)$}{tau}, its angles
with the horizontal \texorpdfstring{$\varphi(t)$}{phi(t)} and its gradients
\texorpdfstring{$\tan(\varphi(t))$}{tan(phi(t))}}

  The central line $(\tau)$, see figure \ref{fig:schq}, is the ``result'' of the
various components (physical time, external and internal rhythms) and it is
supposed to refer to a ``physiological'' time associated to the evolution of the
organism over the course of its life. In order to better understand the
different chronological parts of life, this ``axis'' may be decomposed  in
distinct segments, each being characterized by their angle, $\varphi$,  with
regard to the  abscissas (the $\varphi$ angle under consideration then becomes
that of the tangent), connected by zones with a fast curvature  around specific
times $(t_0, t_1, t_2,\dots)$. We will in particular distinguish five parts
(with unequal lengths).
\begin{enumerate}

\renewcommand{\theenumi}{\Roman{enumi}}
\renewcommand{\labelenumi}{\theenumi}
\item Around $t_{00}$ (which would correspond to the fertilization of the egg that will form the organism or to a mutation which generates a new species), a new segment begins with a very large angle ($80\,^{\circ}$ for example) and consequently with a very high gradient. This segment will correspond to \emph{embryogenesis}.
\item Around $t_0$, there occurs a first  curvature of the axis in order to initiate a segment of which the angle (and the gradient) still remains high (at $60\,^{\circ}$, for example). Time  $t_0$ would correspond to birth\footnote{At germination, for plants.} and the following segment to \emph{growth} (development).
\item Around $t_1$, we would have a new  curvature generating a medium sized angle ($45\,^{\circ}$ for example) with a gradient approaching $1$; $t_1$ would correspond to the apparition of the reproductive faculty (age of puberty\footnote{At the moment of flowering or of fruit-bearing, for plants.}) and to the entering into the phase of \emph{adult maturity}.
\item Around $t_2$, we would have another curvature generating a small angle segment ($30\,^{\circ}$ for example) with a weak gradient; $t_2$ would correspond to the period of loss of fecundity (menopause, eventual andropause)\footnote{At the end of production, for plants.} and to the beginning of \emph{aging} as such.
\item Around $t_3$ the axis becomes horizontal ($\varphi = 0$,  $\tan(\varphi) = 0$) and is definitely broken; $t_3$  represents the moment of \emph{death}.
\end{enumerate}

Concerning the various durations (namely that of the life span $t_3$ - $t_0$), we know by the above mentioned laws of scaling generally encountered in biology, that these durations scale according to the organism approximately by $W_f^{1/4}$,  where  $W_f$  is the mass of the adult organism.

If we now consider $v_t = \tan(\varphi(t))$ as being the ``speed'' of evolution of the physiological time $(\tau)$ with regard to the physical time $t$, we would make the following remarks which motivate the various gradients of $(\tau)$: 
\begin{itemize}
\item between $t_{00}$  and  $t_0$ this speed is very high:  initial cell divisions, morphogenesis, setting in of the first functionalities;
\item between $t_0$ and  $t_1$, the speed remains high; it corresponds to growth, to development, to the completion of the setting in of functionalities, to a high metabolism;
\item between $t_1$ and  $t_2$ the speed is moderate; it corresponds to the regularity of the metabolic reactions, of cellular renewal, etc., that are characteristic of adult age;
\item between $t_2$  and  $t_3$, the speed is low:  lowering of the metabolic rate, of cellular regeneration, of activity; this corresponds to aging;
\item after $t_3$ the speed is null:  it is the death of the organism.
\end{itemize}

\subsection{ The ``helicoidal'' cylinder of revolution \texorpdfstring{$\mathcal{C}_e$\,}{Ce }:  its thread \texorpdfstring{$p_e$}{pe}, its radius \texorpdfstring{$R_i$}{Ri}}

In our qualitative analysis (see \ref{fig:schq}) we have  a cylinder of revolution $\mathcal{C}_e$, with a radius $R_i$, which is winded as a helix having a thread of $p_e$ around the $(\tau)$ axis, without touching this axis but faithfully following its changes of direction.

The thread $p_e$ of this helicoidal cylinder can be assimilated to a period; it corresponds to the \emph{external} cyclical rhythms imposed upon the organism by its environment (annual, lunar, circadian cycles, for instance, see §.\ref{EXT}(EXT)), which are independent physico-chemical rhythms that we have taken into account in the first paragraph; they are essentially of a physical origin and are imposed upon all organisms exposed to them. The $ R_i= 0$ case will be evoked below.

\subsection{ The circular helix \texorpdfstring{$\mathcal{C}_i$}{Ci}  on the cylinder and its thread \texorpdfstring{$p_i$}{pi}}

This circular helix $\mathcal{C}_i$, with a thread $p_i$, is winded around the surface of the cylinder $\mathcal{C}_e$ (it is a ``second order'' helix because the winding cylinder is also helicoidal). We  consider the thread of this helix (which is also a period) to refer to the compacified time $\theta$ (the circle which generates this cylinder) introduced here and associated to the \emph{internal} biological cycles of the organism which are also independent (or almost) from the environment; this is the case, let's recall, for example, of cardiac and respiratory rhythms, of the rhythms of biochemical cascades, etc. (see §.\ref{EXT}(Int)). Let’s also recall that the period associated to these cycles also scale by $W_f^{1/4}$,  at least from  $t_1$  (and also practically from $t_0$).

 To summarize, we thus have, from a biological standpoint, in addition to the objective physical time $t$ (evidently still present and relevant): 
\begin{itemize}
	\item a general temporality of biological evolution $(\tau)$ (the axis);
	\item a temporality associated to the external rhythms (the helicoidal cylinder winded around this axis from a distance) that are characterized by the thread $p_e$;
	\item a temporality associated to the internal rhythms involving a compactification of time:  the helix with a $p_i$ thread at the surface of the cylinder.
\end{itemize}

We should also note that if the radius $R_i$ of the helicoidal cylinder becomes null, it will be reduced to a helix winded around $(\tau)$  and the internal cyclicity will tend to disappear as such (there remains only the external rhythms that are physical). If we may consider the general schema we have presented to concern mainly the properties of the animal world, this last case, where $R_i = 0$, mainly concerns plant. In this sense, the non nullity of $R_i$, that is, the two-dimensionality of the cylindrical surface, should be associated to the greater autonomy --- the rhythms of the  central systems, typically --- and to the autonomous motor capacity which the animal enjoys comparatively to vegetal organisms, the two being obviously correlated. Actually, the rhythms (metabolic, chlorophyllian, of action --- activation of organs\dots) of plants are often completely subordinated to the  physical external rhythms.

 Of course, there is no clear-cut transition, no well-defined boundary between animal and plant life forms in particular in the marine flora/fauna. For this reason, we find the representation of the passing from the one to another in the form of a continuum to be adequate:  the continuous contraction of the helicoidal cylinder which tends towards being a helix, which is a line (the time of plants). The non observability of the difference between animal and plant, in some ``transitional'' cases, would correspond to an interval of biologically possible measurement, with no phase transition (of the type of life form) that is clear or discontinuous. Once the limit, the helicoidal line, is reached, even the three-dimensional embedding space can be collapsed onto the two dimensions:  the rhythm becomes the oscillation of one measurement (of chlorophyllian activity, for example) with regard to the axis of oriented physical time (the spiral is flattened into a sine, for example) as is the case in many 
periodic physical processes.

\subsection{On the interpretation of the ordinate \texorpdfstring{$t'$}{t'}}

Let’s return now to our questioning regarding the interpretation we can give to the ordinate $t'$.  In a certain sense, it is \emph{generated} by the compacified fiber of the temporal rhythms specific to living phenomena. More specifically, it is mathematically necessary as a component of the three-dimensional embedding space of helixes produced by the direct production of the physical time t and of the compacified time $\theta$, which are, according to our hypothesis, two independent dimensions. We already hinted to a possible biological meaning of the $(z)$ coordinate. Then, what could the ordinate $t'$ correspond to, from a biological standpoint?

If we define a speed for the passing of time  $\tau$  comparatively to $t'$ in a way that is similar to the definition of $v_t = \tan(\varphi(t))$,  we will have $v_{t'} = \cotan(\varphi(t))$ ;  at the inverse of $v_t$  (we have  $v_t v_t' = 1$), this velocity is small at first but continues to grow when   $t$ (or  $\tau$)  grows.

In the case where the organism under consideration is the human being,
an interpretation promptly comes to mind. The velocity $v_t$' would correspond
to the \emph{subjective perception} of the speed of the passing of the
``specific'' or physiological time $\tau$:  at first very slow, and then
increasingly rapid with aging. In such case, $t'$ would be the equivalent of a
\emph{subjective time}. One will notice that,
from the quantitative standpoint, if between $t_1$ and $t_2$ (the area of the
adult phase) we confer $\varphi$ with the value of $45\,^{\circ}$ approximately,
as we have already indicated above, the speed of the passing of time $\tau$ with
regard to objective physical time ($v_t$) coincides more or less with the
subjective perception of the passing of this time ($v_{t'}$)  (in fact, 
$\tan(\varphi) \simeq \tan{\varphi} \simeq 1$).  

 As it is matter, here, of human cognitive judgment of the
time flow, we are aware of its historical contingency. The remarks below,
thus, are just informal preliminaries to forthcoming reflections, where the
historicity of young vs. old age perception of time, for example, should be
relativized to specific historical cultures and social frames. We then leave
the reader to have any reflection regarding the subjective perception of time
during youth and old age. We can imagine that such thoughts will coincide with
ours, if we belong to the same ``culture'' (time which passes slowly while young
and, later, very quickly\dots).

In what concerns organisms other than human beings, of which we do not
know  if they have a subjective perception of the speed of the passing of
physiological time $\tau$, it is more difficult to assign a clear status to this
dimension of  $t'$  (although certain relatively evolved species seem likely to
express impatience, for example, or to construct an abstract temporal
representation by exerting faculties of retention and especially of protention).
So would this dimension not begin to acquire a concrete reality only with the
apparition and development of an evolved nervous system (central nervous system,
brain)?  But then what of the bacterium, the amoeba, the paramecium\dots?

Actually, it may be possible to somewhat objectivize the approach by advancing a plausible hypothesis regarding the general character of $t'$:  we could consider that it is a question of a ``temporality'' that is associated to the ``representational'' dimension. Let’s explain.

  Since living organisms are endowed with more or less capacity for retention and \emph{protention} (possibly pre-conscious ``expectation''), we propose (temporarily, this is ongoing work) to base ourselves on the following qualitative argument:  the element of physiological time $d\tau$ is associated to the element of physical time $dt$ and to $dt'$  by the evident relation $d\tau^2 = dt^2 + dt'^2$ ;  it stems from this that $dt'^2$ can be written as $dt'^2 = d\tau^2 - dt^2$  or as
\[dt'^2 = (d\tau -dt)(d\tau +dt)\]

It is then tempting to see in the first factor the minimal expression of
an element of ``retention'' (for physiological time, relatively to physical time)
and in the second the corresponding expression of an element of ``protention''.
The product of the two would generate the temporality component of a
``representation'' which borrows from the ``past'' and from the ``future'', as
constitutive of the flow of biological time. As all living organisms appear to
be endowed with both a capacity for retention --- as rudimentary as it may be
--- and with a protentional faculty (even more rudimentary maybe), the
generality of the dimension $t'$ would be preserved and the ``representational''
capacity (at least in this elementary sense) appears as being a property of
living phenomena, see \cite{longo2011}. This property, for conscious thought,
could even be extended to subjectivity in accordance, in the specific case of
the human being, with the phenomenological analysis with which we began:  $dt'^2$
would be a form, as elementary as infinitesimal, of the ``extended present'', in
the husserlian tradition, described by other analyzes, such as the coupling of
oscillators in \cite{varela1999}.

Finally, it would be the two-dimensionality $t \times t'$ --- (physical time) $\times$ (representation time) --- which would enable to \emph{mark out} the temporality of living phenomena, which may be represented in the geometrical way as we have described in this paper.

\nocite{bailly1993}
\nocite{Bak88}
\nocite{bourgine2004}
\nocite{botzung2008}
\nocite{chaline1999}
\nocite{edelman2001}
\nocite{edelman2000}
\nocite{feynman1967}
\nocite{vangelder1999}
\nocite{horsfield1977}
\nocite{lecointre2001}
\nocite{levanquyen2003}
\nocite{nottale1993}
\nocite{peters1986}
\nocite{petitot2003}
\nocite{rosen2005}
\nocite{schmidtnielsen1984}
\nocite{szpunar2007}
\nocite{varela1999}
\nocite{vaz1978}
\nocite{waddington1981}
\nocite{west1997}
\nocite{west1999}
\nocite{leme98}

\bibliographystyle{alpha}
        \bibliography{bib}

\markboth{Bibliographie}{Bibliographie}

\end{document}